\documentclass[runningheads]{llncs}
\usepackage{graphicx}
\usepackage[hyphenbreaks]{breakurl}
\usepackage[hyphens]{url}
\usepackage{moreverb,url}

\usepackage[colorlinks,bookmarksopen,bookmarksnumbered,citecolor=black,urlcolor=black]{hyperref}

\newcommand\BibTeX{{\rmfamily B\kern-.05em \textsc{i\kern-.025em b}\kern-.08em
		T\kern-.1667em\lower.7ex\hbox{E}\kern-.125emX}}

\usepackage{graphicx}

\usepackage{smartdiagram}
\smartdiagramset{set color list={
		gray!30,
		gray!30,
		gray!30,
		gray!30,
		gray!30,
		gray!30,
	},
	sequence item border color=black,
	sequence item font size=16,
}

\usepackage{float}

\usepackage{color}
\usepackage{colortbl}

\usepackage{amssymb}
\usepackage{comment}
\usepackage[multiple]{footmisc}
\usepackage{amsmath}

\usepackage{lipsum}
\usepackage{pfnote}
\usepackage{multirow}

\usepackage{pgfplots}
\usepackage{tikz}
\usetikzlibrary{arrows,decorations.pathmorphing,fit,positioning}
\usepackage{xcolor}
\usetikzlibrary{patterns}
\usepackage{amstext}

\usetikzlibrary{mindmap,trees}

\usepackage{fixfoot} 
\DeclareFixedFootnote{\myfnone}{one}  

\usepackage{enumitem}

\usepackage{colortbl}

\usepackage{bbding} 

\definecolor{maroon}{cmyk}{0,0.87,0.68,0.32}

\usetikzlibrary{positioning,calc}

\usepackage{tikz}
\usetikzlibrary{trees}

\tikzset{level 1/.style={level distance=2cm, sibling distance=10cm}}
\tikzset{level 2/.style={level distance=2cm, sibling distance=3cm}}

\tikzset{bag/.style={text width=10em, text centered,yshift=-0.5cm}}

\usepackage{smartdiagram}
\usesmartdiagramlibrary{additions}

\usetikzlibrary{positioning,calc}

\usetikzlibrary{arrows,shapes,positioning,shadows,trees}

\tikzset{
	basic/.style  = {draw, text width=4cm, drop shadow, font=\sffamily, rectangle},
	root/.style   = {basic, rounded corners=2pt, thin, align=center,
		fill=green!30},
	level 2/.style = {basic, rounded corners=6pt, thin,align=center, fill=green!60,
		text width=2em},
	level 3/.style = {basic, thin, align=left, fill=pink!60, text width=2em}
}

\def\addlegendimage{\csname pgfplots@addlegendimage\endcsname}

\begin{document}
\title{Computational Analysis of Insurance Complaints: GEICO Case Study }
%
%
\author{Amir Karami\inst{1} \and
Noelle M. Pendergraft\inst{2}}

\authorrunning{Karami et al.}
%
\institute{College of Information and Communications, University of South Carolina, USA \\
	\email{karami@sc.edu}\\ \and South Carolina Honors College, University of South Carolina, USA \\ \email{noellep@email.sc.edu}	}
\maketitle              
\begin{abstract}
The online environment has provided a great opportunity for insurance policyholders to share their complaints with respect to different services. These complaints can reveal valuable information for insurance companies who seek to improve their services; however, analyzing a huge number of online complaints is a complicated task for human and must involve computational methods to create an efficient process. This research proposes a computational approach to characterize the major topics of a large number of online complaints. Our approach is based on using topic modeling approach to disclose the latent semantic of complaints. The proposed approach deployed on thousands of GEICO negative reviews. Analyzing 1,371 GEICO complaints indicates that there are 30 major complains in four categories: (1) customer service, (2) insurance coverage, paperwork, policy, and reports, (3) legal issues, and (4) costs, estimates, and payments. This research approach can be used in other applications to explore a large number of reviews. 

\end{abstract}
\section{Introduction}

Consumers post millions of their opinions for products and services on the website hosting online reviews. These hug amount of online reviews play an important role for business growth \cite{he2017examining}. While positive feedback can help business growth, negative feedback or complains have negative impacts on business revenue along with higher hiring cost. In the \$175 billion US auto insurance industry, different websites have been developed for collecting the thousands of valuable reviews of policyholders. This data scale precludes manual annotation and organization, and is a motivation for applying computational approaches. 

There are two types of data: structured and unstructured. While structured data such as numbers is clearly defined, unstructured data such as text is the opposite. There are also two types of methods for computational data analysis: supervised and unsupervised. The former one needs training data such as classification and regression methods, but the latter one doesn't need training data such as clustering methods \cite{8215705}. Some supervised methods have been used on insurance-related structured data such as investigating customer retention and insurance claim patterns using variables such as postal code, vehicle age, age, and gender \cite{smith2000analysis}.

Surveys and polls are the traditional method used to gain insight into understanding of customer opinions regarding services or products \cite{chang2009national}.  However, with advent of web, several studies have been developed to use online reviews to collect public opinion data with respect to companies \cite{cheung2012drives} and showed a positive correlation between the average rating of online reviews and sales \cite{archak2011deriving}. 

Online review analysis has been used to identify important factors of customer satisfaction in different business applications such as hotel industry \cite{li2013determinants} and restaurants \cite{yan2015customer}. Insurance-related studies also analyzed online reviews such as using several variables such as policyholder age, review rating, and postal code to predict whether a customer renews or retains her/his policy and to detect high risk customers \cite{smith2000analysis}. Although the related studies provided insightful views to the literature, there isn't any study to explore insurance-related online reviews using unstructured data and unvised methods. 

Among several insurances companies, GEICO is one of the top insurance companies with more than 16 million policies and 24 million insured vehicles\footnote{https://www.geico.com/about/corporate/at-a-glance/}. This company has thousands of online reviews\footnote{https://clearsurance.com/best-car-insurance-2018} that manually analyzing them is beyond human capabilities. This paper utilized an un approach to analyze unstructured online reviews. We applied our approach on more than 1000 GEICO online reviews to identify and categorize major complaints. Not only insurance companies and providers but also other organizations can use this study to better understand the concerns of their consumers and reflect upon those accordingly.

\section{Methodology and Results}

\subsection{Data Collection}
 We collected this research data from format from \url{https://www.consumeraffairs.com/insurance/geico.htm} . This website provides a platform for verified reviewers to share their opinions with respect to different services and products. Each review contained a text review and star rating from 1 to 5. To focus on complaints, we considered 1 and 2 star vehicle insurance reviews and found 1,371 negative reviews. This dataset is available at \url{https://github.com/amir-karami/Geico_Negative_Reviews}.

\subsection{N-gram Analysis}
In this step, we applied bigram analysis. N-gram approach analyzes the sequence of n words in a given corpus such as unigram analysis, n-gram of size 1, bigram analysis, n-gram of size 2, and trigram analysis, n-gram of size 3. This step came with removing the words that don't have any semantic value like “the” and “a” using a standard stopwords list.  The n-gram analysis represents some themes behind the complaints such as ``customer service," ``body shop repairmen," ``police report," ``auto policy," ``claim adjuster," ``payments," and ``rental coverage".  We also tracked the frequency of GEICO's competitors in the data and observed that Progressive, AllState, and State Farm are the top three high frequency companies mentioned by the users.  It seems that users compared GEICO with these companies in the reviews.

\subsection{Topic Discovery and Analysis}
\begin{figure}[H]
	
	\begin{center}
		\scalebox{0.6}{
\includegraphics[scale=0.55]{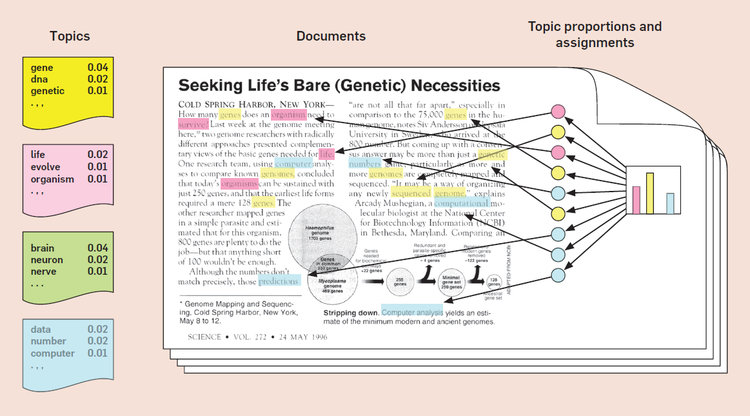}
		}
		\caption{Intuition Behind LDA \cite{blei2012probabilistic}}
		\label{fig:ldaintuition}
	\end{center}	
\end{figure}

 Although n-gram analysis provided insightful information about the complaints, some combinations like ``red light," ``parking lot," and ``phone calls" didn't provide meaningful patterns. Therefore, we need to consider not only the sequence of words but also the overall semantic in our corpus. To disclose another semantic layer in our data, we looked at advanced text mining methods. There are two text mining approaches: supervised and unsupervised. For detecting interesting patterns, supervised techniques need training data, but unsupervised techniques don't require any manual effort \cite{karami2017fuzzy,karami2015fuzzy}. Due to the large number of complaints without having training data or manual label, we selected the unsupervised approach in this research. 
Among different unsupervised text mining methods, probabilistic topic modeling helps the topic discovery \cite{karami2015flatm,karami2015fuzzyiconf}. While several topic models has been developed for different applications, latent Dirichlet allocation (LDA) is the most widely-used topic model that has been employed in a wide ranges of applications such as libraries \cite{collins2018social}, health \cite{karami2018characterizing,webb2018characterizing}, politics \cite{karami2018mining}, and spam detection \cite{karami2015online,karami2014improving,karami2014exploiting}. This model assigns the words that are semantically related to each other in a topic represented a theme \cite{blei2003latent}.  For example in a corpus, LDA assigns ``\textit{gene}," ``\textit{dna}," and ``\textit{genetic}" to a topic representing ``genetic" theme (Fig \ref{fig:ldaintuition}). LDA assumes that each topic is a distribution over words and each document is a mixture of topics.

We used the MALLET implementation\footnote{http://mallet.cs.umass.edu/topics.php} of LDA to find the topics of complaints \cite{karami2014fftm}.  The complaints were imported in the MALLET format along with removing the stopwords. To estimate the number of topics, we used the log-likelihood analysis method to find the number of topics associated with the highest log-likelihood \cite{shaw2017computational}. The log-likelihood estimation indicated that the number of topics at 30 was the optimum point for soft clustering of the words.  Table 2 shows the topic number and the detected words for each topic. Each labels was selected based on the overall theme in a topic. For instance, ``Rental Car and Deductible" was selected as the label of words in topic 1 including ``\textit{rental, claim, pay, days, and deductible}".

\begin{table}[htp!]
	\scriptsize
	\caption{Topics of Complaints}
	\label{tab:topics}
	
	\begin{tabular}{|p{0.5cm}|p{6.8cm}|p{5.cm}|}
		
		\hline
		\rowcolor[gray]{.9} \textbf{T\#}  &  \textbf{Topic} & \textbf{Label}\\ \hline
		T1 &  rental claim pay days deductible claims collision damage coverage cover  & Rental Car \& Deductible\\ \hline
		T2 &	claim file person claims run find fraud accident complaint investigator &	Fraud Investigation\\ \hline
		T3 &	quote policy agent online rate change price called higher rep &	Higher Price Than Quoted\\ \hline
		T4 &	policy called payment paid bill asked cancelled received monthly payments &	Billed After Cancellation\\
		\hline
		T5 &	called claim adjuster received left number contact weeks agent service &	Numerous Calls for Claim Adjustment\\ \hline
		T6 &	company people business money owner insured practices action unfair lawsuit law &	Lawsuit for Unfair Business Practice\\ \hline
		T7 &	customer service representative gave customers mistake continue loyal spoke poor &	Poor Customer Service Attitude Toward Loyal Customer \\ \hline
		T8 &	shop repair body adjuster parts damage estimate damaged replaced refused & Repair Payment and Adjuster\\ \hline
		T9&	police fault report accident light red turn stopped stated damage &	Police Fault Report of an Accident \\ \hline
		T10&	customer service company experience worst horrible terrible rude cheap treated &	Poor Customer Service\\ \hline
		T11&	total loss dollars worth totaled adjuster price offered sell cash &	Estimate for Totaled Vehicle\\ \hline
		T12&	damage front bumper door accident side hit lot parking paint rear left& 	Damage in Parking Lot \\ \hline
		T13&	coverage full pay money bought customer liability save service options &	Coverage Options\\ \hline
		T14&	policy spoke explained good needed representative clear issues information complete &	Clear and Complete Policy/Info \\ \hline
		T15&	check called back weeks send paperwork months fax wrong forward &	Slow Paperwork\\ \hline
		T16&	information stated auto representative number provided party companies requested correct &	Auto Representative\\ \hline
		T17&	left attorney street office deal client completely lie wrong showed &	Court \\ \hline
		T18&	information claim weeks called contacted case response woman handle uninsured &	Uninsured Case \\ \hline
		T19&	back pay money day long wait hours making calls working &	Long Waiting for Money Back\\ \hline
		T20&	asked called back supervisor wanted manager rude rep speak customer &	Rude Employees \\ \hline
		T21&	truck back fix drive collision problem month ago water flood &	Flood Damage \\ \hline
		T22&	accident medical pay pain bills totaled injuries hospital horrible injury &	Medical Costs\\ \hline
		T23&	account payment bank money card credit refund make debit checking &	Payments\\ \hline
		T24&	covered roadside assistance tire service problem towing called tires gas &	Roadside Assistance\\ \hline
		T25&	letter received state send coverage email mail dmv phone address &	Contact with DMV\\ \hline
		T26&	paid money stolen thing job months paying totaled life hard &	Payments in Hardship Situations\\ \hline
		T27&	policy son daughter driver added family driving wife coverage husband &	Add Family Member to Policy\\ \hline
		T28&	claim accident adjuster client made fault case liability filed denied &	Refuse to Pay a Claim\\ \hline
		T29&	years months record driving past give switched insure great high &	Not Considering Long-Term Driving Record\\ \hline
		T30&	rates premium company policy increase year cost times increased raised &	Rate Increase \\ \hline
		
	\end{tabular}
\end{table}

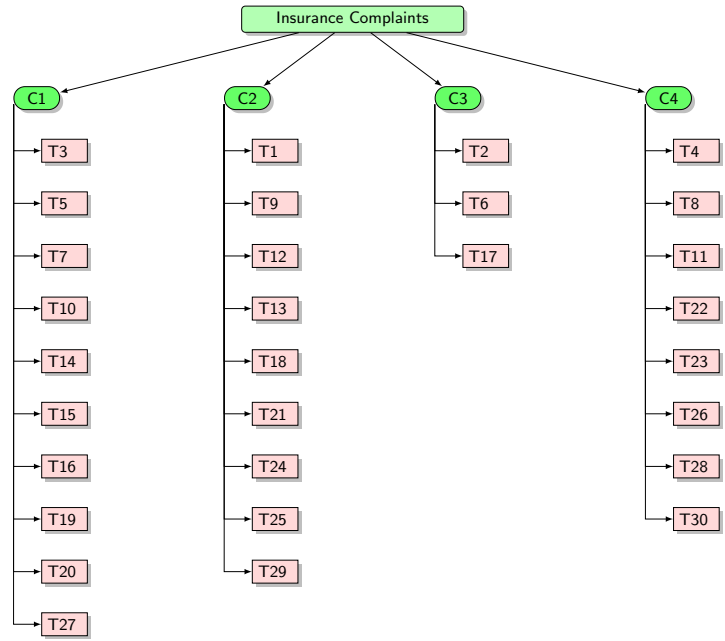
\begin{figure}[htp!]

\begin{center}
	\scalebox{0.7}{
\begin{tikzpicture}[scale=1,
level 1/.style={sibling distance=40mm},
edge from parent/.style={->,draw},
>=latex]

\node[root] {Insurance Complaints}
child {node[level 2] (c1) {C1}}
child {node[level 2] (c2) {C2}}
child {node[level 2] (c3) {C3}}
child {node[level 2] (c4) {C4}};

\begin{scope}[every node/.style={level 3}]
\node [below of = c1, xshift=15pt] (c11) {T3};
\node [below of = c11] (c12) {T5};
\node [below of = c12] (c13) {T7};
\node [below of = c13] (c14) {T10};
\node [below of = c14] (c15) {T14};
\node [below of = c15] (c16) {T15};
\node [below of = c16] (c17) {T16};
\node [below of = c17] (c18) {T19};
\node [below of = c18] (c19) {T20};
\node [below of = c19] (c111) {T27};

\node [below of = c2, xshift=15pt] (c21) {T1};
\node [below of = c21] (c22) {T9};
\node [below of = c22] (c23) {T12};
\node [below of = c23] (c24) {T13};
\node [below of = c24] (c25) {T18};
\node [below of = c25] (c26) {T21};
\node [below of = c26] (c27) {T24};
\node [below of = c27] (c28) {T25};
\node [below of = c28] (c29) {T29};

\node [below of = c3, xshift=15pt] (c31) {T2};
\node [below of = c31] (c32) {T6};
\node [below of = c32] (c33) {T17};

\node [below of = c4, xshift=15pt] (c41) {T4};
\node [below of = c41] (c42) {T8};
\node [below of = c42] (c43) {T11};
\node [below of = c43] (c44) {T22};
\node [below of = c44] (c45) {T23};
\node [below of = c45] (c46) {T26};
\node [below of = c46] (c47) {T28};
\node [below of = c47] (c48) {T30};
\end{scope}

\foreach \value in {1,2,3,4,5,6,7,8,9,11}
\draw[->] (c1.195) |- (c1\value.west);

\foreach \value in {1,...,9}
\draw[->] (c2.195) |- (c2\value.west);

\foreach \value in {1,...,3}
\draw[->] (c3.195) |- (c3\value.west);

\foreach \value in {1,...,8}
\draw[->] (c4.195) |- (c4\value.west);
\end{tikzpicture} 
}
\caption{Categories of Complaints}
\label{fig:inssurancecat}
\end{center}	
\end{figure}

After the topic discovery step, we annotated (labeled) 30 topics (Table \ref{tab:topics}).  To reach an agreement on the labels, the researchers compared notes and discussed differences in their topics until an agreement was reached. Based on the labels, we identified four categories for the annotated topics: Customer Service Related Complaints (C1), Insurance Coverage, Paper Work, Policy, and Report Related Complaints (C2), Legal Issues Related Complaints (C3), and Costs, Estimates, and Payments Related Complaints (C4) (Fig \ref{fig:inssurancecat}).

The first category covers the topics related to the situations in which customers contact customer service to ask a question(s), fix a problem(s), or get help /information /quote, such as calling for claim adjustment in topic 5, but they weren't happy or satisfied after this process. This category includes inappropriate behavior of GEICO employees like in topics 7, 10, and 20. The second category is related to insurance coverage directly, such as rental car service and deductible issue in topic 1 and roadside assistant in topic 24, or indirectly, like police fault report of an accident in topic 9. The next category includes the legal cases such as the fraud detection in topic 2. Finally, the last category represents the complaints related to money, including payments, such as charging a customer account after policy cancellation in topic 4, estimates, such as giving low estimates for a totaled vehicle in topic 11, and costs ,such as medical costs. The map of topics on the four categories indicates that more than 60\% of topics are related to C1 and C2. This means that GEICO needs to invest more on customer service and insurance coverage, paperwork, policy, and report to improve the customer satisfaction.

\section{Conclusion}

Online review websites provide a platform for people to share their opinions on different products and services. Previous studies showed that online reviews are important for people and have direct impact on the growth and revenue of insurance companies. Understanding the topics of complaints in the reviews can help the insurance companies to improve their products and services. However, analyzing a huge number of reviews is not an easy task for human. 

This paper provides a computational content analysis approach to collect, analyze, and visualize unstructured text data. Our research methodology decodes complaints in online reviews with low rating and can be applied to other business applications.  Our results showed that most of the topics were related to customer service. The results from this study have practical implications for numerous stakeholders that are in insurance and other industries, providing customer information to managers. Our future research studies will include the complains of other insurance companies, such as AllState, Statefarm, and Progressive, along with considering time and location variables. 

\section*{Acknowledgements}   
This research is supported in part by the South Carolina Honors College Science Undergraduate Research Fellowship. All opinions, findings, conclusions, and recommendations in this paper are those of the authors and do not necessarily reflect the views of the funding agency.

%
%
%
\bibliographystyle{splncs04}
%
\bibliography{refrence}
\end{document}